\documentclass[aps,prb,twocolumn]{revtex4}

\usepackage{txfonts}
\usepackage{epsfig}

\begin{document}

\title{Location of crossings in the Floquet
spectrum of a driven two-level system.}
\author{C.E.~Creffield}
\affiliation{Instituto de Ciencia de Materiales (CSIC), Cantoblanco,
E-28049, Madrid, Spain}
\thanks{Current address: Dipartimento di Fisica, Universit\`a di Roma
``La Sapienza'', Piazzale Aldo Moro 2, I-00185, Roma, Italy.}

\date{\today}

\begin{abstract}
Calculation of the Floquet quasi-energies of a system
driven by a time-periodic field is an efficient
way to understand its dynamics. In particular, the
phenomenon of dynamical localization can be related to the presence
of close approaches between quasi-energies (either crossings
or avoided crossings). We consider here a driven two-level system
and study how the locations of crossings in
the quasi-energy spectrum alter as the field parameters are changed.
A perturbational scheme provides a direct connection between the
form of the driving field and the quasi-energies which is exact
in the limit of high frequencies. We firstly obtain relations for the
quasi-energies for some common types of applied field in the
high-frequency limit, and then show
how the locations of the crossings drift as the frequency is reduced.
We find a simple empirical formula which
describes this drift extremely well in general, and which we conjecture
is exact for the specific case of square-wave driving.
\end{abstract}


\maketitle

\section{Introduction}

The two-level system is a simple model which has
been applied to a great variety of physical problems. One
application of growing importance is that of quantum computing
\cite{divincenzo}, since any quantum two-level system has the potential
to act as a quantum bit. For this reason the coherent control of quantum
states in these systems has recently become the focus of intense
investigation \cite{vion, cole}. A concrete example of such a two-level system
is provided by a particle tunneling between two potential wells, which
can be experimentally realized by confining an electron to a pair of
coupled quantum dots \cite{klitzing}.
One method of controlling such a system, without destroying its coherence,
is to apply oscillatory electric fields \cite{leo_dots}. Such fields
are able to produce the phenomenon known as {\em coherent destruction of
tunneling} (CDT), in which the tunneling of the particle is suppressed
when the parameters of the field are tuned to various ``magic'' values.
As the applied field is time-periodic, Floquet analysis \cite{shirley}
has been applied to explain this non-intuitive result, and it has been
shown \cite{hanggi_1, hanggi_2} that CDT is closely related to the
presence of crossings or avoided crossings in the spectrum of Floquet
quasi-energies.

The driving field most frequently considered is of sinusoidal
form, and studies using
CDT as a means of quantum-control have generally concentrated on
varying either the envelope \cite{holthaus_pulse} or frequency
\cite{holthaus_chirp} of a sinusoidal signal. In this work,
however, we instead consider the effect of altering the signal's
{\em waveform}. By using a perturbational method we first show how the
waveform can be directly related to the quasi-energy spectrum, and give
analytic results for sinusoidal, square-wave and triangular waveforms.
These results are precise in the limit of high frequency. As the
frequency is reduced, however, the locations of the crossings drift
away from these values. This effect is extremely difficult to treat
analytically, and such efforts \cite{barata,frasca,delgado}
produce complicated results which are difficult to interpret.
Empirically, however, we find a simple formula which describes the drifting
with good accuracy for many waveforms, and appears to be {\em exact} for
the case of the square-wave. We thus provide a means for predicting the
locations of quasi-energy crossings for a given driving field in both high
and low frequency regimes.

\section{Method}

\subsection{Model Hamiltonian}

We consider a charged particle confined to a double quantum-dot system,
described by the Hamiltonian:
\begin{equation}
H = {\tilde t} \left( c_L^{\dagger} c_R^{ } + h.c. \right) +
\left( E_L(t) n_L + E_R(t) n_R \right)
\label{ham_tunn}
\end{equation}
where the subscript $L/R$ denotes the left/right quantum dot,
$c_j^{\dagger}$ and $c_j^{ }$ are creation and annihilation operators for
a particle in dot $j$, and $n_j =c_j^{\dagger}c_j^{ } $ is the usual number
operator. The tunneling between the two dots is described by the hopping
parameter $\tilde t$, and $E_j(t)$ is the electrical potential of
the external driving field. Clearly only the potential
difference between the two dots is physically of importance, and
so we can use the symmetric parameterization:
\begin{equation}
E_L = \frac{E}{2} f(t), \quad  E_R = -\frac{E}{2} f(t)
\end{equation}
where $E$ is the potential of the driving field and $f(t)$ is
a $T$-periodic function describing its waveform.
The Hamiltonian (\ref{ham_tunn}) has been written using a basis of
{\em localized} states, but it may be easily transformed to the
standard two-level form via a SU(2) rotation, yielding the result:
\begin{equation}
H = \frac{\Delta}{2} \sigma_z \ + \ \frac{E}{2} f(t) \sigma_x,
\label{ham_2lev}
\end{equation}
where $\sigma_i$ are the standard Pauli matrices. In this representation
the basis states used are {\em extended} states, formed by symmetric
and anti-symmetric combinations of the localized states. In the absence
of a driving field ($E=0$) it is clear that the two eigenstates of this
Hamiltonian consist of a symmetric ground state, and an excited
anti-symmetric state. The splitting between these two levels is
given by $\Delta$, which is related to the inter-dot tunneling via
$\Delta = 2 {\tilde t}$.

\subsection{\label{proceed} Floquet theory}

As the function $f(t)$ is periodic in time, the Floquet theorem may
be used to write solutions of the time-dependent Schr\"odinger equation
as $\psi(t) = \exp[-i \epsilon_j t] \phi_j(t)$, where $\phi_j(t)$ is
a function with the same periodicity as $f(t)$ and is called a
Floquet state, and $\epsilon_j$ is termed the quasi-energy.
Although the Floquet states have an explicit
time-dependence, their periodicity means that the
dynamics of the system on time-scales much larger than the period of
the driving field is effectively given {\em only} by the quasi-energies.
In particular, if the two quasi-energies approach degeneracy, the
dynamics of the system on this time-scale will appear to be frozen,
producing the effect of CDT. Consequently, determining the quasi-energies
provides a simple and direct way of studying the long time-scale
behavior of the system, and indicates whether CDT can occur \cite{caveat}.
In this work we restrict our attention to driving functions
which possess the symmetry $f(t) = - f(t + T/2)$.
Imposing this restriction means that the Hamiltonian (\ref{ham_tunn})
is invariant under the generalized parity operation
$x \rightarrow -x, t \rightarrow t + T/2$, and as a consequence
the two Floquet states will also possess this symmetry,
one being even and the other being odd. The von Neumann-Wigner theorem
\cite{neumann} thus allows the two quasi-energies to cross
as an external parameter, such as the field
strength, is varied. Breaking this symmetry by choosing an alternative
form for the driving field would mean that the quasi-energies would be
forbidden to cross, and thus close
approaches between the quasi-energies could only consist of avoided
crossings.

The Floquet states and their quasi-energies may be conveniently
obtained from the eigenvalue equation:
\begin{equation}
\left[ H(t) - i \frac{\partial}{\partial t} \right] \phi_j(t)
= \epsilon_j \phi_j(t) .
\label{floquet}
\end{equation}
To obtain approximate solutions to this equation
we follow a perturbation scheme introduced
originally by Holthaus \cite{holthaus} to treat both the two-level system
and driven superlattices, and which was generalized recently to also
include the effects of inter-particle interactions \cite{creffield}.
In this approach
the Hamiltonian (\ref{ham_tunn}) is divided into two parts: $H_t$
which contains the tunneling terms, and $H_I$ which holds
the electric field terms. We then find the eigensystem of the operator
${\cal H}_I(t) = H_I - i {\partial}/{\partial t}$ by working in
an {\em extended} Hilbert space of time-periodic functions \cite{sambe},
and apply the tunneling Hamiltonian as a perturbation. A consequence of
dividing the Hamiltonian in this way is that the perturbation theory works
well in the high-frequency limit $\omega \gg {\tilde t}$,
but breaks down in the opposite limit when the tunneling provides
the dominant energy-scale of the problem \cite{hanggi_2}.

For the Hamiltonian given in Eq.\ref{ham_tunn}, the problem of finding
the eigensystem of ${\cal H}_I(t)$ simply requires the solution of
two uncoupled differential equations:
\begin{eqnarray}
\left( -\frac{E}{2} f(t) - i \frac{d}{dt} \right) \phi_+(t) &=& \epsilon_+
\phi_+(t), \\
\left( \frac{E}{2} f(t) - i \frac{d}{dt} \right) \phi_-(t) &=& \epsilon_-
\phi_-(t).
\end{eqnarray}
These can be integrated immediately, giving the solutions:
\begin{eqnarray}
\phi_{\pm}(t) &=& \exp[\pm i E F(t)/2] \exp[i \epsilon_{\pm} t], \\
\mbox{where} \ \ F(t) &=& \int_0^{t} f(t') dt' .
\label{sols}
\end{eqnarray}
The periodicity of the Floquet states clearly
requires that $\epsilon_{\pm} = 0 \ \mbox{mod} \ \omega$.
Without loss of generality we can restrict the quasi-energies to lie in
the ``first Brillouin zone'' ($- \omega/2 \leq \epsilon < \omega / 2$),
and thus to lowest order in the perturbation theory
they are degenerate and zero. Standard degenerate perturbation theory 
can now be used to evaluate the first-order correction to the quasi-energies,
requiring only that we work in the the extended Hilbert space of
$T$-periodic functions by defining an appropriate scalar product:
\begin{equation}
\langle \langle \phi_m | \phi_n \rangle \rangle_T =
\frac{1}{T} \int_0^{T} \langle \phi_m(t') | \phi_n(t') \rangle dt'
\end{equation}
where $\langle \cdot | \cdot \rangle$ is the usual scalar product
for the spatial component of the wavefunctions,
and $\langle \cdot | \cdot \rangle_T$ denotes the integration over
the compact time coordinate.

As the tunneling component of the Hamiltonian $H_t$ is acting
as the perturbation, the first-order approximation to the
quasi-energies is given by the eigenvalues of the perturbing matrix:
\begin{equation}
\langle \langle H_t \rangle \rangle_T =
\pmatrix{ &0 \ & {\tilde t} \ \langle \phi_-^2 \rangle_T \cr
	  & {\tilde t} \ \langle \phi_+^2 \rangle_T \ & 0}
\label{effective}
\end{equation}
Comparing this expression with the original tunneling Hamiltonian
(\ref{ham_tunn}) reveals that the action of the applied field is to
renormalize the tunneling terms by the factors
$\langle \phi_{\pm}^2 \rangle_T$. As $\phi_+$ is the complex
conjugate of $\phi_-$, the quasi-energies take the simple form:
\begin{equation}
\epsilon_{\pm} = \pm \frac{\Delta}{2}
\left| \langle \phi_+^2 \rangle_T \right| ,
\label{quasi}
\end{equation}
where
\begin{equation}
\langle \phi_+^2 \rangle_T = \frac{1}{T} \int_0^T \exp[i E F(t)] dt
\label{central}
\end{equation}
and $F(t)$ is defined in Eq.\ref{sols}. Clearly the quasi-energies
can only become degenerate when they are both equal to zero, and
we can note from Eq.\ref{effective} that this corresponds, as expected,
to the destruction of the effective tunneling.

\section{Results}

To obtain the Floquet quasi-energies for comparison with the
prediction of Eq.\ref{central}, the numerical technique
described in Ref.\cite{creffield} was used. This involves
evaluating the unitary evolution operator for one period
of the field $U(T,0)$ and obtaining its eigenvalues, which
are related to the quasi-energies via $\lambda_j = \exp[-i \epsilon_j T]$.
Using this method to obtain the quasi-energies,
a standard bisection algorithm could then be used to find the location
of the quasi-energy crossings to a high degree of accuracy.

The dynamical behavior of the system was also examined directly by
integrating it over long time-periods, with the particle initially
located in the left quantum dot.
To quantify to what extent the tunneling between the
left and right quantum dots was destroyed, the
probability that the particle was in the left quantum dot ($P_L(t)$)
was measured throughout the time evolution.
We denote the minimum value of $P_L$ attained during this period
to be the ``localization'', and thus high values of localization
correspond to the presence of CDT, while low values reveal that
the particle is able to tunnel from one side to the other, and
is therefore delocalized.

\subsection{Sinusoidal driving}

\begin{figure}
\centerline{\epsfxsize=60mm \epsfbox{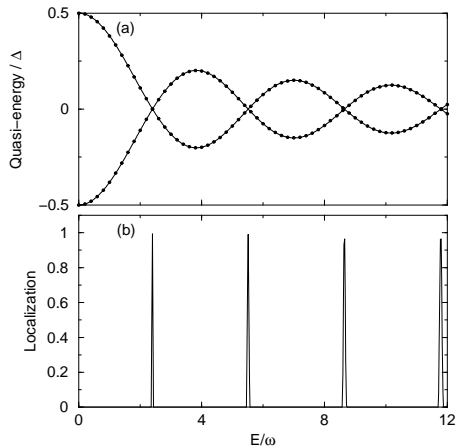}}
\caption{(a) Quasi-energies for a sinusoidal driving field,
of frequency $\omega=8$. Circles
indicate exact results, lines the perturbative result
$\pm (\Delta/2) J_0(E/\omega)$.
(b) Localization in the driven system. Spikes
in the localization are centered on crossings of the quasi-energies.}
\label{j0}
\end{figure}

We begin with the most familiar case, when the driving field
has the form $f(t) = \cos \omega t$. The procedure
outlined in Section \ref{proceed} can be followed straightforwardly,
leading to the result that:
\begin{equation}
\langle \phi_{+}^2 \rangle_T = \frac{1}{T}
\int_0^T \exp[i E \sin (\omega t) / \omega] \ dt .
\end{equation}
By making use of the standard identity:
\begin{equation}
\exp[i E \sin (\omega t) / \omega] = \sum_{m=-\infty}^{\infty}
J_m \left( E / \omega \right) \exp[i m \omega t]
\end{equation}
this expression can be substantially simplified, yielding the final
result that $\epsilon_{\pm} = \pm (\Delta / 2) \ J_0(E / \omega)$.
This reproduces the well-known result that for sinusoidal
driving CDT occurs when the ratio of the field strength to its
frequency is equal to a root of the Bessel function $J_0$.
In Fig.\ref{j0}a the locations of the quasi-energies are shown for
a fixed frequency $\omega = 8$ as a function of $E/\omega$. It can
be seen that the perturbative result works extremely well in this regime
(high-frequency). Fig.\ref{j0}b shows the localization produced
by the field, as defined above. As expected, at the points where
the quasi-energies cross the tunneling dynamics of the system is blocked,
producing sharp spikes in the localization, centered on the crossings.

\begin{figure}
\centerline{\epsfxsize=60mm \epsfbox{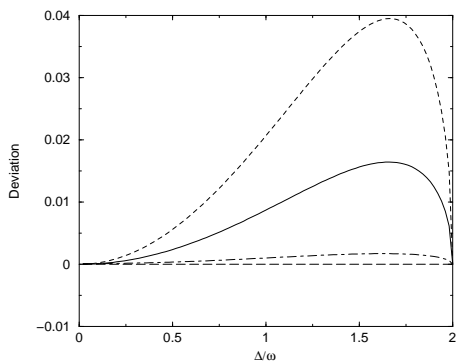}}
\caption{Deviation of the first crossing-manifold from the empirical
fitting function. Solid line indicates sinusoidal driving;
dashed line = square-wave driving;
dotted line = triangular driving.
For the square-wave the deviation is smaller than $10^{-7}$.
The dot-dashed line gives the deviation for the Fourier
expansion of the square-wave, truncated at two terms.}
\label{deviation}
\end{figure}

To investigate how the crossings move away from these
points as the driving frequency is reduced, their locations are shown
as a function of $1/\omega$ in Fig.\ref{drifts}a. In accordance with the
von Neumann-Wigner theorem \cite{neumann, hanggi_2} we can readily see
that the set of crossings form one-dimensional manifolds. As
$\omega$ tends to infinity the crossings occur at the roots of $J_0$,
as predicted by the perturbation theory,
and this remains a good approximation for frequencies as low as
$\omega = \Delta$. Below this value, however, the crossings smoothly
drift away from these locations, and evolve towards the points
$\Delta / \omega = 2 n$ (where $n$ is a positive integer),
as was seen earlier in Ref.\cite{hanggi_2}.
This limiting behavior in the low-frequency regime
was also predicted in Ref.\cite{ulloa},
where a similar pattern of crossing-drift was observed in an investigation
of a related model. The form of Fig.\ref{drifts}a immediately suggests
fitting the manifolds of crossings with quadrants of ellipses:
\begin{equation}
\left( \frac{E / \omega}{y_n} \right)^2 +
\left( \frac{\Delta / \omega}{2 n} \right)^2 = 1 ,
\label{fitting}
\end{equation}
where $y_n$ is the $n$-th root of $J_0(y)$. It can be seen
in Fig.\ref{deviation} that this
simple parameterization fits the results extremely well for the
first crossing-manifold, and that the difference between the exact
location of the crossing and the fitting function 
($E_{fit}/\omega - E_{exact}/\omega$)
never exceeds a value of $0.02$. The degree of deviation becomes larger
as the order of the crossing increases, but
nonetheless is only visible in Fig.\ref{drifts}a
for the fourth and fifth crossing-manifolds.

\begin{widetext}
\begin{center}
\begin{figure}
\centerline{\epsfxsize=160mm \epsfbox{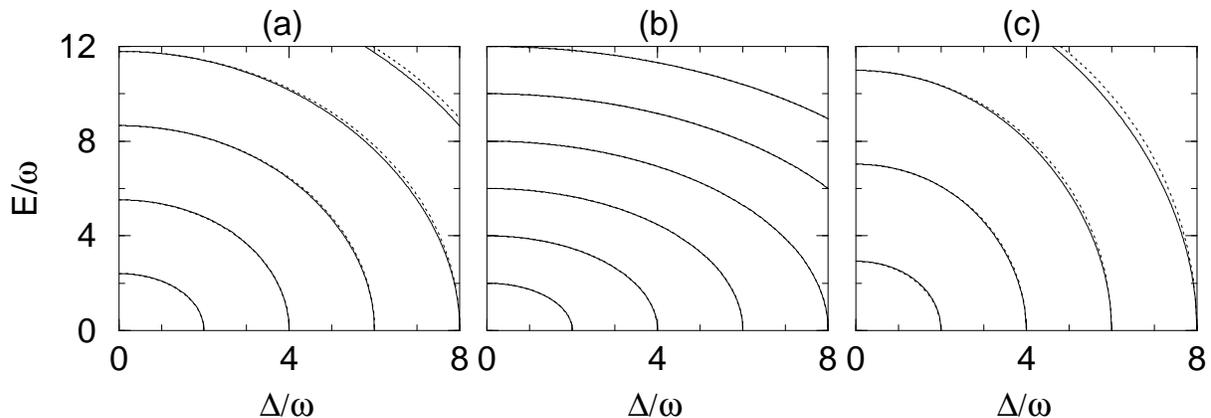}}
\caption{Location of crossings of quasi-energies, in
each case the crossings fall on one-dimensional manifolds.
(a) sinusoidal driving; (b) square-wave driving; (c) triangular driving.
Dotted lines indicate the empirical fitting function Eq.\ref{fitting}.}
\label{drifts}
\end{figure}
\end{center}
\end{widetext}

\begin{figure}
\centerline{\epsfxsize=60mm \epsfbox{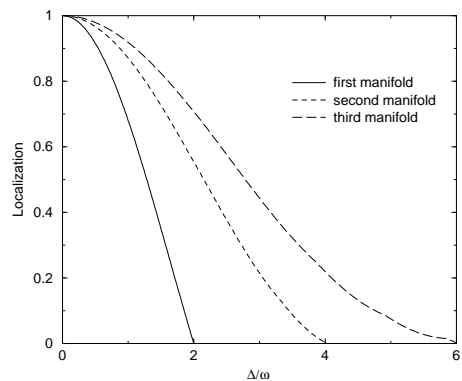}}
\caption{Localization produced by a sinusoidal field, for
$(E, \omega)$ coordinates lying on the first three crossing-manifolds.}
\label{arc}
\end{figure}

In Fig.\ref{arc} the localization is plotted as $\omega$ is reduced
from a high value towards zero, with $E$ set to hold the ratio
$E/\omega$ on a crossing-manifold. For each point the
system was evolved over 200 periods of the driving field to study
how effectively the field could maintain a localized state. For the
high-frequency regime, $\omega \geq \Delta$, the localization is excellent
at all the crossings, with less than 0.1 of the particle density leaking
across to the right-side dot during the time-evolution.
As can be expected, the high-order crossings,
which occur at higher values of $E$, can maintain better
levels of localization than the low-order crossings \cite{hanggi_2}.
This difference becomes more pronounced as the frequency is reduced,
and although the localization in all cases decays smoothly to zero,
the localization at the higher-order crossings decays much more slowly.
For frequencies as low as $\omega=0.4 \Delta$, however,
the inhibiting effect of CDT is still evident for all the crossings,
indicating that even low-frequency fields may serve a useful role
in stabilizing electron-leakage from quantum dot devices.

\subsection{Square-wave driving}

Square-wave driving has been considered to a lesser extent than
the sinusoidal case, although it is also an easily realizable
waveform in experiment. Ref.\cite{zhao_square} investigated the
case of a superlattice driven by a square-wave field, and found that
for suitable choices of parameters CDT would indeed occur, while
sinusoidal driving of this system could only produce partial
CDT \cite{zhao_bloch}. Recently in Ref.\cite{dignam} it has been
shown that in a superlattice CDT can only be produced if the crossings
of the quasi-energies are {\em equally} spaced, which clearly does
not occur for sinusoidal driving. For this reason it is
of interest to derive the behavior of the quasi-energies for
square-wave driving to see explicitly how this condition is fulfilled.

\begin{figure}
\centerline{\epsfxsize=60mm \epsfbox{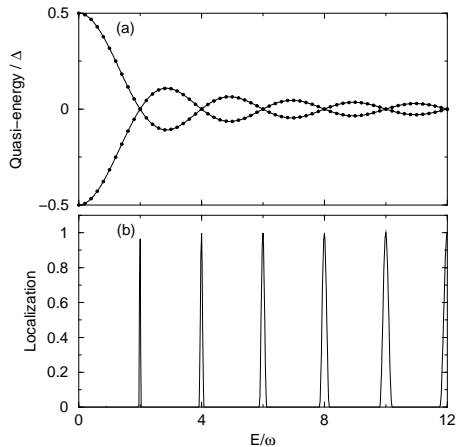}}
\caption{(a) Quasi-energies for a square-wave driving field,
of frequency $\omega=8$. Circles
indicate exact results, lines the perturbative result Eq.\ref{pert_sq}.
(b) Localization produced by the driving field.}
\label{sinc}
\end{figure}

We consider the square-wave driving field,
$f(t) = \Theta(t) - 2 \Theta(t - T/2)$, defined over the interval
$0 \leq t < T$. The integrations required to obtain the
quasi-energies may again be done straightforwardly, giving the
result that:
\begin{equation}
\epsilon_{\pm} = \pm \frac{\Delta}{2} \
\frac{ \sin \left( \pi E / 2 \omega \right) }{\pi E / 2 \omega} .
\label{pert_sq}
\end{equation}
From this it is immediately clear that the crossings
are equally spaced as required, being given by
the condition $E / \omega = 2 n$ where $n$ is a positive integer.
In Fig.\ref{sinc}a the quasi-energies obtained for a frequency of
$\omega=8$ are shown in comparison with the above result,
and it can be clearly seen that the agreement is excellent. Below this
figure is plotted the localization produced by the field, and as
for the case of sinusoidal driving, the crossings of the quasi-energies
correspond to sharp spikes in the localization, verifying that CDT
is indeed occurring.

In Fig.\ref{drifts}b the drifting of the crossings as the frequency
is reduced is shown. The behavior is strikingly similar
to that observed for sinusoidal driving, and
accordingly we use the same functional form (\ref{fitting})
to fit the crossing-manifolds, with the y-intersections
now given by $y_n = 2 n$. The fit is so good that on this
plot no differences can be seen between the exact results
and the fits. This is corroborated by Fig.\ref{deviation},
where the deviation from the exact result for the lowest manifold
can be seen to be negligible in comparison with the sinusoidal case,
and within the accuracy of the numerical procedures
the fit is {\em identical} with the exact result. We therefore conjecture
that this fitting is, in fact, exact for the case of square-wave
driving. We also show on this plot the result obtained for a
bandwidth-limited square-wave, obtained by truncating its Fourier expansion
at two terms. We see that the addition of just the second
term to the sinusoidal driving already reduces the deviation of the fit
from the exact result considerably. Truncating the series at higher
points produces steady improvements in the fit, strongly supporting
the conjecture that the fit is exact when all terms are included.

\subsection{Triangular driving}

\begin{figure}
\centerline{\epsfxsize=60mm \epsfbox{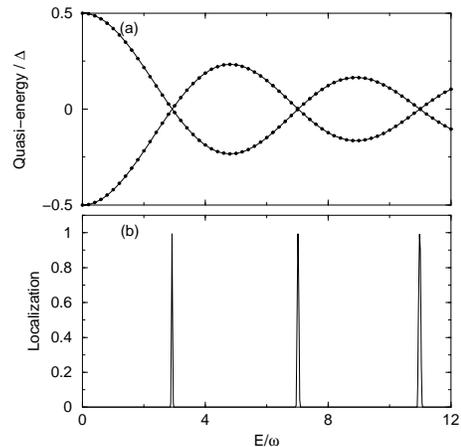}}
\caption{(a) Quasi-energies for a triangular driving field,
of frequency $\omega=8$. Circles
indicate exact results, lines the perturbative result Eq.\ref{monster}.
(b) Localization produced by the driving field.}
\label{tri}
\end{figure}

We now consider another easily obtainable form of driving, the
triangular waveform:
\begin{equation}
f(t) = \left\{
\begin{array}{r@{\quad \quad}l}
1 - 4 t / T & \mbox{for} \ 0 \leq t \leq T/2 \\
-3 + 4 t/T &  \quad \ T/2 < t \leq T .
\end{array}
\right.
\end{equation}
For this case a closed form solution can again be obtained for
the behavior of the quasi-energies, involving the Fresnel
sine and cosine functions, $S(x)$ and $C(x)$. The full expression
for the quasi-energies is given by:
\begin{equation}
\epsilon_{\pm} = \frac{\Delta}{\sqrt{2 x}} \left[
\cos(x \pi/4) C(\sqrt{x/2}) + \sin(x \pi/4) S(\sqrt{x/2}) \right]
\label{monster}
\end{equation}
where $x = E / \omega$. In Fig.\ref{tri} it can be seen that this
function is indeed an excellent approximation to the true quasi-energies,
and that CDT again occurs at the points of quasi-energy crossings. The
roots of Eq.\ref{monster} may be found numerically,
yielding the result that the first three crossings occur when
$E/\omega = 2.92519, \ 7.02525$ and 10.9864. Observing the behavior
of the Fresnel functions \cite{abram} reveals that for $x > 1$ they both
make small amplitude, decaying oscillations about a value of 0.5, which
allows the condition for crossings to be written in the simpler, though
approximate, form $\tan (x \pi/4) \simeq -1$. The crossing condition therefore
reduces to the simple result $E / \omega \simeq 4 n - 1$, as
may be seen from the exact values given above, which
becomes increasingly accurate for larger values of $E / \omega$.

In Fig.\ref{drifts}c it can be seen that the crossing-manifolds
for this form of driving have a similar elliptical form to the
previous cases. Using the same fitting function (\ref{fitting})
as before, with the y-intercepts given by the roots of
Eq.\ref{monster}, gives an accurate description of their behavior,
as may be seen in Fig.\ref{deviation}. Although the fit is not
as good as for the sinusoidal case, the maximum deviation is still
less than $0.04$. As seen previously, the fit is best for
the lowest-order manifolds, with small deviations being visible in
the higher-order manifolds. Nonetheless, in all cases the
fitting function gives an impressively accurate approximation
to the true result.

\section{Conclusions}

In summary, it has been shown how changing the waveform of a periodic
driving-field can be used to modify the location of the quasi-energies
of a two-level system. A procedure has been given which
relates the waveform explicitly to the quasi-energy spectrum,
allowing the positions of the quasi-energy crossings to be located exactly
in the limit of high frequency. For various driving fields, including
the cases we consider here, an analytic form can be obtained for
the quasi-energies, and in other cases they may be obtained
numerically with little difficulty. This gives the prospect of
designing the waveform to create a desired behavior of the quasi-energy
spectrum in a direct and straightforward way.

It has also been shown how the positions of the quasi-energy
crossings drift as the frequency is reduced from the
high-frequency limit. For the driving
fields considered here, the crossings fall approximately
onto elliptical manifolds,
and for the case of square-wave driving it appears that this description
is exact. We have examined this behavior for many other waveforms,
and we conclude that this form of the crossing-manifolds is very general.
Using the perturbation theory to find the crossings in the
high-frequency limit, and then making use of this drifting
behavior, allows the positions of the quasi-energy crossing to be accurately
located in {\em all} regimes of driving. This gives more flexibility
in experiment, as the high-field regime may either be difficult
to attain, or may induce undesirable transitions to higher energy levels,
breaking the two-level approximation. Although the degree of localization
that the field can maintain is reduced in the low-frequency regime,
it can still produce a useful reduction of the leakage from quantum
dot devices, and thereby enhance their decoherence time, which has
many possible applications to the coherent control of mesoscopic systems.

\acknowledgments
This research was supported by the EU through the
TMR programme ``Quantum Electron Transport in the
Frequency and Time Domains''. The author 
thanks Gloria Platero for discussions, and acknowledges the hospitality
of the International Institute for Applied Systems Analysis (IIASA)
in Vienna, where part of this work was carried out.


\begin{thebibliography}{}
\bibitem{caveat}
{It should be noted, however, that although quasi-energy
degeneracy is necessary to produce CDT
it is not sufficient. For example, if the Floquet
states themselves have a large amplitude of oscillation, the particle
will not be localized on short time-scales, and thus CDT will not occur.}
\end{thebibliography}
\end{document}